\documentclass[12pt]{iopart}

\newcommand{\slpar}{\partial\!\!\!/}
\usepackage{graphicx}  
\begin{document}

\title{The nonrelativistic limit of the Majorana equation and its simulation in trapped ions}

\author{L Lamata$^1$, J Casanova$^1$, I L Egusquiza$^2$, and E Solano$^{1,3}$}

\address{$^1$ Departamento de Qu\'{\i}mica F\'{\i}sica, Universidad del Pa\'{\i}s Vasco -- Euskal Herriko Unibertsitatea, Apartado 644, 48080 Bilbao, Spain}
\address{$^2$ Departamento de F\'{\i}sica Te\'orica, Universidad del Pa\'{\i}s Vasco -- Euskal Herriko Unibertsitatea, Apartado 644, 48080 Bilbao, Spain}
\address{$^3$ IKERBASQUE, Basque Foundation for Science, Alameda Urquijo 36, 48011 Bilbao,
Spain}

\begin{abstract}
We analyze the Majorana equation in the limit where the particle is at rest. We show that several counterintuitive features, absent in the rest limit of the Dirac equation, do appear. Among them, Dirac-like positive energy solutions that turn into negative energy ones by free evolution, or nonstandard oscillations and interference between real and imaginary spinor components for complex solutions. We also study the ultrarelativistic limit, showing that the Majorana and Dirac equations mutually converge. Furthermore, we propose a physical implementation in trapped ions.
\end{abstract}

\maketitle

\section{Introduction}
Quantum simulators are quantum controllable devices aiming at the reproduction of complex quantum system dynamics \cite{Feynman82}. Among their promised advantages are an exponential speedup in the computational time with respect to classical computers for simulating quantum systems, and the possibility to analyze systems difficult to experiment with in the lab \cite{Lloyd96,Buluta09}. Many physical systems have already been emulated in a quantum simulator. Among them, spin models \cite{Friedenauer08,Kim10}, quantum phase transitions \cite{Greiner02}, quantum chemistry \cite{Lanyon10}, spin statistics \cite{OBrien}, and relativistic systems \cite{Lamata07, Gerritsma1,Casanova1,Szameit,Gerritsma2,Lamata11,CasanovaQFT,Weitz}. Recently, a theoretical proposal for the quantum simulation of the Majorana equation and unphysical operations in trapped ions was proposed \cite{Casanova2}.

The Majorana equation is a non-Hamiltonian relativistic wave equation that takes the form \cite{Majorana37}
\begin{equation}
i\hbar\slpar\psi=mc\psi_c,
\end{equation}
where $\slpar=\gamma^\mu\partial_\mu$ and $\gamma^\mu$ are the Dirac matrices~\cite{Thaller}, and $\psi_c$ is the charge conjugate of $\psi$. More explicitly, for 1+1 dimensions, the Majorana equation is written as
\begin{equation}
\label{eq:major11}
i\hbar\gamma^{0}\partial_0\psi + i\hbar\gamma^{1}\partial_{x}\psi = mc \psi_{c},
\end{equation}
where the $2 \times 2$ matrices
\begin{equation}
\gamma^{0}=\left(
\begin{array}{cc}
1&0\\
0&-1
\end{array}\right), \ \
\gamma^{1}=\left(
\begin{array}{cc}
0&1\\
-1&0
\end{array}\right),
\end{equation}
and
\begin{equation}
\psi_{c} = \gamma^{1}\gamma^{0}\psi^{*} .
\end{equation}
Notice that, as opposed to the Dirac equation, this is a non-Hamiltonian equation due to the simultaneous presence of $\psi$ and $\psi_c$. Thus, this equation contains counterintuitive features, like the nonexistence of energy eigenstates. In Ref.~\cite{Casanova2}, it was shown that the Majorana equation can be mapped to a higher-dimensional Dirac equation that has a Hamiltonian evolution, such that the Majorana dynamics may be implemented in a controllable quantum optics setup as trapped ions~\cite{Casanova2}.

In this work, we go a step further and analyze the strict nonrelativistic limit in which the Majorana particle is at rest. Thus, in Section~\ref{SectionRestLimit} we consider the limit $p/mc\ll 1$, where all moments of the particle momentum $p$ are negligible. We will show that, surprisingly, novel features emerge, even in the rest limit of the particle $p\simeq0$: if we begin with a Dirac-like spinor with positive energy, the Majorana equation will mix its dynamics with the equivalent to the Dirac equation negative energy branch, and it will produce nonstandard interference between real and imaginary components. Thus, the rest limit of the Majorana equation behaves in a totally different way to the equivalent limit of the Dirac equation. Indeed, this is the limit where both equations are more different. In Section~\ref{SectionIonSimul}, we show how to implement this dynamics in a trapped-ion system and, in Section~\ref{SectionConclusions}, we present our conclusions.  Finally, for the sake of completeness, we include in the Appendix the exact solution of the Dirac and Majorana equations and its mutual convergence in the ultrarelativistic limit in which $mc\ll p$.

\section{The rest limit of the 1+1 Majorana equation\label{SectionRestLimit}}

The dynamics of a general spinor for the 1+1-dimension Majorana equation, in the limit in which $p=0$ (i.e., $p\ll mc$), can be obtained in the following way:

In this limit~(\ref{eq:major11}) is equivalent to 
\begin{equation}
\partial_t\psi=-\frac{mc^2}{\hbar}\sigma_y\psi^*\,,
\end{equation}
where \(\sigma_y\) is the Pauli matrix. Notice that the complex conjugate of \(\sigma_y\) is \(-\sigma_y\), so this entails
\begin{equation}
\partial_t\psi^*=\frac{mc^2}{\hbar}\sigma_y\psi\,,
\end{equation}
and as a consequence
\begin{equation}
\partial^2_t\psi=-\left(\frac{mc^2}{\hbar}\right)^2\psi\,.
\end{equation}
The solution must therefore be of the form
\begin{equation}
\psi_M(t)=\cos\omega t \,\psi_M(0)+\sin\omega t\,\xi\,,
\end{equation}
with \(\omega=mc^2/\hbar\), and \(\xi\) is fixed by consistency to \(-\sigma_y\psi^*_M(0)\), thus arriving at
\begin{equation}\label{MajoranaSolut}
\psi_M(t)=\cos\omega t \,\psi_M(0)-\sin\omega t\,\sigma_y\psi^*_M(0)\,.
\end{equation}
This is to be compared to the corresponding rest mass frame Dirac equation, which is given in this case by 
\begin{equation}
\partial_t\psi_D=-i\omega\sigma_z\psi_D\,,
\end{equation}
whence the solution directly reads
\begin{equation}
\psi_D(t)=\exp\left(-i\omega t\sigma_z\right)\psi_D(0)=\cos\omega t\,\psi_D(0)-i\sin\omega t\,\sigma_z\psi_D(0)\,.\label{DiracSolut}
\end{equation}
We can now compare both solutions, giving
\begin{equation}
\psi_M(t)=\frac{1}{2}\left[\psi_D(t)+\psi_D(-t)\right]-\frac{1}{2}\sigma_x\left[\psi^*_D(t)-\psi^*_D(-t)\right]\,.
\end{equation}
 
Accordingly, what we obtain is that the rest dynamics of the Majorana equation is quite more involved than the one for Dirac equation. For Dirac, the positive and negative energy branches do not mix in this limit, and the only dynamics is a global phase associated to the rest energy of the Dirac particle (positive or negative depending on the branch). For Majorana, the dynamics contains one term that is like the Dirac dynamics, but other three novel terms: in the first one, the temporal dynamics is like in the former one, but reversed (evolution backwards in time). In the second one, there is a complex conjugation of the spinor, and a spinor flip: in the Dirac language, positive and negative energy components would get mixed by this dynamics; in the Majorana equation this is more subtle, given that there are not energy eigenstates associated to this equation and positive and negative energy branches do not have here a clear meaning. In the third one, there is a simultaneous time-reversed dynamics, spin flip, and complex conjugation.

Summarizing, what we obtain is that even such a naively simple limit as the rest limit contains intriguing features for the case of the Majorana equation: interference between forward- and backward-time evolving solutions, between real and complex spinor components inside the same Dirac-like positive or negative branch, and between spinors associated to these branches.

\begin{figure}[t] \centering
\includegraphics[width=0.6\linewidth]{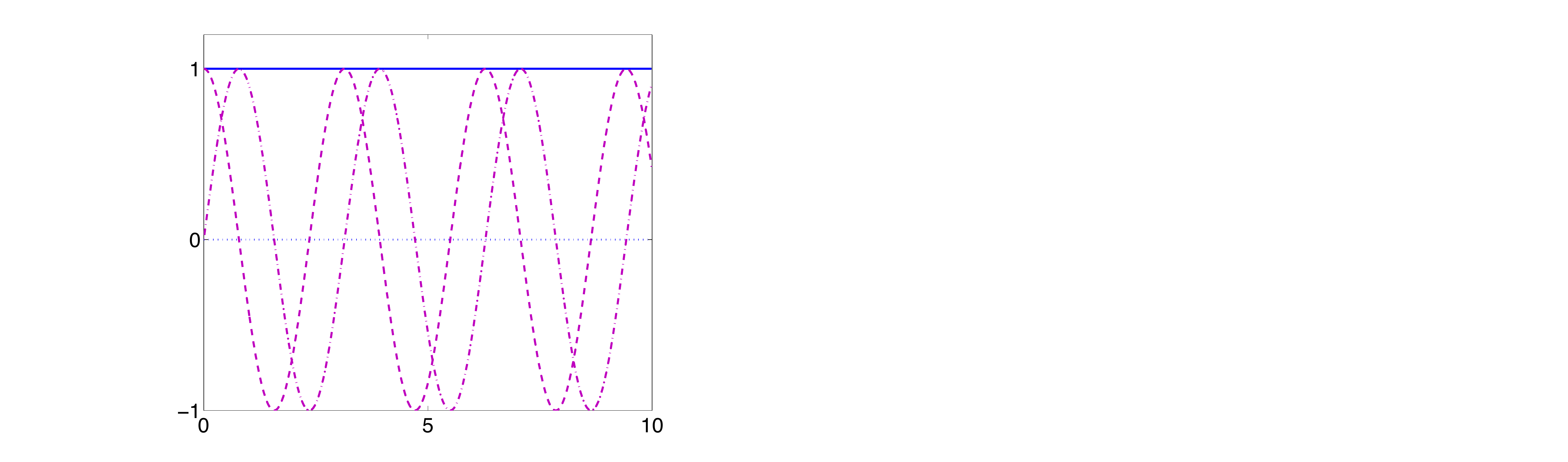}
\caption{$\langle \psi(t)|\sigma_z|\psi(t)\rangle$ for states $|\psi(0)\rangle=(1,0)^T$ (Dirac: solid; Majorana: dashed), and $|\psi(0)\rangle=(1,i)^T/\sqrt{2}$ (Dirac: dotted; Majorana: dotted-dashed).}
             \label{Fig1}
\end{figure}

In Fig. \ref{Fig1} we plot the average value of $\sigma_z$, $\langle \psi(t)|\sigma_z|\psi(t)\rangle$, for states $|\psi(0)\rangle=(1,0)^T$ (Dirac: solid; Majorana: dashed), and $|\psi(0)\rangle=(1,i)^T/\sqrt{2}$ (Dirac: dotted; Majorana: dotted-dashed). It can be appreciated that, despite the fact that the Dirac dynamics is trivial (the average value does not change with time in either case), the Majorana case is not: there are oscillations in $\langle \psi(t)|\sigma_z|\psi(t)\rangle$ even for a particle at rest, and even in the case in which initially only the upper component of the spinor is present.

This is clearly obtained with Eqs. (\ref{MajoranaSolut}) and (\ref{DiracSolut})
\begin{equation}
\langle\sigma_z\rangle_D(t)=\psi^{\dag}_D(t)\sigma_z\psi_D(t)=\psi^{\dag}_D(0)\sigma_z\psi_D(0)\,,
\end{equation}
since \(\sigma_z\) is proportional to the Dirac Hamiltonian; while for Majorana we have
\begin{equation}
\!\!\!\!\!\!\!\!\!\!\!\!\!\! \psi_M^{\dag}(t)\sigma_z\psi_M(t)=\cos(2\omega t)\psi_M^{\dag}(0)\sigma_z\psi_M(0)-\sin(2\omega t) \mathrm{Im}\left[\psi_M^{\dag}(0)\sigma_x\psi_M^*(0)\right],
\end{equation}
reproducing the behaviour in Fig. \ref{Fig1}.

\section{Quantum simulation in trapped ions\label{SectionIonSimul}}
In order to perform a quantum simulation of the rest limit of the Majorana equation in trapped ions, one should point out that the operation $\psi\rightarrow K\psi=\psi^*$ is unphysical. In order to implement the dynamics of Eq. (\ref{MajoranaSolut}), we follow Ref.~\cite{Casanova2}:
We perform a mapping from the two-component Majorana complex spinor onto a four component real spinor,
\begin{equation}
\psi_M(t)=\left(\begin{array}{c}\psi_1^r+i\psi_1^i\\\psi_2^r+i\psi_2^i\end{array}\right)\in C_2\rightarrow\psi_4=\left(\begin{array}{c}\psi_1^r\\\psi_2^r\\\psi_1^i\\\psi_2^i\end{array}\right)\in R_4.
\end{equation}
Accordingly, the rest-limit Majorana equation in the doubled space takes the form
\begin{equation}
i\hbar\partial_t\psi_4=-mc^2\sigma_x\otimes\sigma_y\psi_4.
\end{equation}
This equation, as the one with $p\neq 0$, is straightforwardly simulated with a four-level trapped ion or two two-level ions, as shown in Ref.~\cite{Casanova2}. In the rest case considered here, no coupling to the motion is needed, and just a two-spin Hamiltonian should be necessary. In order to measure an observable $A$ in the Majorana space, one should measure an observable $M^\dag A M$ in the doubled space, where $M=(I\,\, iI)$ and $M \psi_4 = \psi_M$. In order to reproduce the dynamics from Fig.~\ref{Fig1}, one should measure the four-level population by resonance fluorescence with a cyclic transition, following standard electron-shelving techniques~\cite{IonReview}.

\section{Conclusions\label{SectionConclusions}}
We have analyzed the strict nonrelativistic limit of the Majorana equation in which the particle is at rest. We have shown that intriguing phenomena appear, even for this rest situation, as compared to the Dirac equation: Dirac-like positive energy components that turn into negative ones, or real and complex spinor components that interfere in a nontrivial way. These novel features can be straightforwardly simulated with two trapped ions.

\section*{Acknowledgements}
 L. L. thanks the European Commission for a Marie Curie IEF grant. J. C. acknowledges support from Basque Government grant BFI08.211. I. L. E. is grateful to Basque Government grant IT559-10. E. S. acknowledges funding from Basque Government grant IT472-10, Spanish MICINN FIS2009-12773-C02-01, SOLID, and CCQED European projects.

\section*{Appendix}
From the comparison of the Majorana and Dirac equations it is intuitive that the ultrarelativistic limit, that is, $m\to0$, will coincide for both. The detailed computation, however, is instructive. Both equations presented here are exactly solvable; the traditional technique for the Dirac case is to notice that the solutions of the Dirac equation are also solutions of the Klein-Gordon equation, and use separation of variables. We shall present a rather subtle proof that a similar result holds for the Majorana equation; this will allow us to assert an ansatz for the solution of the Majorana equation that will provide us with the exact solution. We will then take the limit $m\to0$, and will derive in which way Majorana solutions tend to Dirac solutions.

Let us write the Majorana equation as
\begin{equation}
i\hbar\partial_t\psi=-i\hbar c \sigma_x\partial_x\psi-i m c^2\sigma_y\psi^*\,,
\end{equation}
and use Fourier transform,
\begin{equation}
\psi(x,t)=\int_{-\infty}^{+\infty}\frac{\mathrm{d}p}{\sqrt{2\pi\hbar}}\psi_p(t)e^{i p x/\hbar}\,.
\end{equation}
It follows that
\begin{equation}
i\hbar\partial_t\psi_p=c p \sigma_x\psi_p-i m c^2\sigma_y\psi_{-p}^*\,.
\end{equation}
Notice that the complex term couples the $p$ and $-p$ components.

Directly from the previous expression, by taking the complex conjugate and interchanging at the same time $p$ and $-p$, we infer
\begin{equation}
-i\hbar\partial_t\psi_{-p}^*=-cp\sigma_x\psi_{-p}^*-i m c^2 \sigma_y\psi_p\,.
\end{equation}
In this derivation it is relevant to notice that $\sigma_y$ is purely imaginary, so $i \sigma_y$ is real.
On the other hand, we can solve $\psi_p^*$ from the expression above, as
\begin{equation}
\psi_{-p}^*=-\frac{\hbar}{mc^2}\sigma_y\partial_t\psi_p-\frac{p}{mc}\sigma_z\psi_p\,,
\end{equation}
and on differentiating, we obtain
\begin{equation}
-i\hbar\partial_t\psi_{-p}^*=\frac{i\hbar p}{m c}\sigma_z\partial_t\psi_p+\frac{i\hbar^2}{ mc^2}\sigma_y\partial_t^2\psi_p\,.
\end{equation}
Putting all together, we conclude that the components $\psi_p$ obey the Klein-Gordon equation
\begin{equation}
\hbar^2\partial_t^2\psi_p+\left(p^2c^2+m^2c^4\right)\psi_p=0\,.
\end{equation}

It follows that $\psi_p(t)$ will have the solution
\begin{equation}
\psi_p(t)=\cos\left(\omega_p t\right)\psi_p(0)+\frac{\sin\left(\omega_p t\right)}{\omega_p}\dot{\psi}_p(0)\,,
\end{equation}
where $\omega_p=\sqrt{p^2 c^2+m^2 c^4}/\hbar$, and using the equation above, 
\begin{equation}
\dot{\psi}_p(0)=-\frac{i c p}{\hbar}\sigma_x\psi_p(0)-\frac{m c^2}{\hbar}\sigma_y\psi_{-p}^*(0)\,,
\end{equation}
so
\begin{equation}
\psi_p(t)=\cos\left(\omega_p t\right)\psi_p(0)-\frac{\sin\left(\omega_p t\right)}{\hbar\omega_p}\left[i c p \sigma_x\psi_p(0)+ m c^2\sigma_y\psi_{-p}^*(0)\right]\,.
\end{equation}
The limit $p\gg mc$ now makes sense, and gives (together with a condition over the time, $t\ll2\hbar|p|/m^2 c^3$)
\begin{equation}
\psi_p(t)\approx \exp(-i c p t \sigma_x/\hbar)\psi_p(0)\,,
\end{equation}
which coincides with the aproximate solution in the same limit for the Dirac equation. The exact solution for the Dirac equation in all the range of values of $p$ is actually
\begin{equation}
\psi^D_p(t)=\cos\left(\omega_p t\right)\psi^D_p(0)-i \frac{\sin\left(\omega_p t\right)}{\hbar\omega_p}\left[cp \sigma_x+m c^2 \sigma_z\right]\psi^D_p(0)\,,
\end{equation}
which allows for direct comparison with the Majorana solution. For example, clearly when $m=0$ both solutions coincide.

\end{document}